**Thermal decay rate of a metastable state with two degrees of freedom:
dynamical modeling versus approximate analytical formula**

I. I. Gontchar[a] and M. V. Chushnyakova[b,c,*]

*Physics and Chemistry Department, Omsk State Transport University 644046, Omsk, Marxa 35, Russia*
*Physics Department, Omsk State Technical University 644050, Omsk, Mira 11, Russia*
*Department of Applied Physics, Tomsk Polytechnic University, 634034, Tomsk, Lenina 30, Russia*

*Corresponding author.
e-mail:* maria.chushnyakova@gmail.com (M.V. Chushnyakova)

Accuracy of the Kramers approximate formula for the thermal decay rate of the metastable state is studied for the two-dimensional potential pocket. This is done by the comparison with the quasistationary rate resulting from the dynamical modeling. It is shown that the Kramers rate is in agreement with the quasistationary rate within the statistical errors provided the absorptive border is far enough from the potential ridge restricting the metastable state. As the absorptive border (or its part) gets closer to the ridge the Kramers formula underestimate the quasistationary rate. The difference reaches approximately the factor of 2 when the absorptive border coincides with the ridge.

Metastable system; quasistationary decay rate; Kramers rate; dynamical modeling

PACS numbers: 05.60.-k, 82.20.Pm

**1. Introduction**

One encounters the problem of thermal decay of a metastable (quasistationary) state in many branches of physics, chemistry and biology: fission of excited nuclei and dissociation of a molecule are just two examples (see e.g. Refs. [1-8]). Let us note, that presently more than 10% of electricity on Earth is produced by the nuclear power plants using the thermal nuclear fission process which is not completely understood theoretically [4, 9, 10].

The rate of thermal decay of a metastable (quasistationary) state in the presence of friction is evaluated for one-dimensional motion using the formulas derived by Kramers in Ref. [11] or their modifications [12-15]. The rates calculated according to those formulas are expected to agree with the long time limit of the escape rate obtained using either the stochastic differential equations (the Langevin equations) or the corresponding partial differential equations (the Fokker-Planck equation or the Smoluchowski equation). Below this limit is referred to as the Quasistationary Dynamical Rate (QDR) and denoted as $R_D$.

The problem of agreement between the approximate analytical rates (Kramers rates, $R_K$) and $R_D$ (which is supposed to be exact within the statistical and numerical errors) was studied in several works [13-18] for the case of one degree of freedom (1D). In particular, in [13-16] the applicability of the Kramers formula for the case of high temperature was studied. In [12, 15] corrections to the Kramers formula were proposed which account for the anharmonicity of the potential. In [13, 14] the concept of the mean first passage time (developed long time ago in [19, 20]) was applied for the microcanonical ensemble. In [14], for the first time by our knowledge, the dependence of the QDR upon the position of the absorptive border was studied for the harmonic potential (i.e. the potential constructed of two smoothly joined parabolas). It was shown in that work that when the absorptive border is far away from the barrier point, the Kramers formula agrees with the QDR, whereas as the border gets closer to the barrier the QDR exceeds the Kramers rate significantly (see Figs. 1 and 2 in Ref. [14]). In Ref. [13], from the concept of the mean first passage time, a correction to the Kramers formula was obtained that accounts for the final distance between the barrier and absorptive points (see Eq. (23) in that work). This correction goes over to the factor 2 when these two points coincide. Let us note that the consideration in Refs. [13-18] was restricted to the case of the overdamped motion.

In Refs. [21-23] the Kramers formula was generalized for the case of several degrees of freedom. Every now and again these generalizations were confronted with results of numerical modeling [23-25]. However, to our knowledge, a systematic comparison for the multidimensional problem is absent in the literature. In particular, it is not studied whether the influence of the absorptive border in the 2D case is the same, as for the 1D case. The aim of the present work is to remedy the situation to certain extent.

In his pioneering work [11] Kramers indicated the atomic nuclear fission process as one of the three physical problems for which his results might be applied. Presently vast literature exists on the application of the Kramers approach to nuclear fission (see e.g. Refs. [3, 4, 13, 21, 23-30]). Our work stems from the nuclear fission problem too. Therefore we devote special attention to the cases with typical dependences of the potential upon the collective coordinates. We also take into account the "distance" between the position of the potential barrier and the absorptive border. However, the results obtained might be of general interest since many features of the model (e.g. the canonical ensemble, the coordinate-independent friction and inertia parameters, the single-barrier potential) are common in different applications of the Brownian motion.

The paper is organized as follows. Sec. 2 is devoted to the description of the model. Approximate analytical formula as well as the recipe for calculation of the quasistationary dynamical decay rate are discussed here. In Sec. 3 we compare the Kramers rate with the QDR for different potential landscapes and different layouts of the absorptive border. The friction strength is also varied. In Sec. 4 we summarize our results. In Appendix A we discuss the details of finding the QDR and its errors. In Appendix B the table comprising the parameters of the modeling is presented.

## 2. The model

### 2.1. The scenario

The motion of the Brownian particle is characterized by two collective coordinates $q_0$ and $q_1$ which we consider to be dimensionless. In the case of nuclear fission, $q_0$ is responsible for the elongation of the fissioning nucleus and $q_1$ represents the necking of the shape. Fig. 1 illustrates the relation between the coordinates and the shape (only mirror-symmetric fission is considered). Note, that the odd shapes like in the upper left and lower right corners of this figure possess extremely large potential energies and therefore are never reached during the modeling. Fig. 2 represents the potential landscape along which the Brownian particle moves. This is a realistic potential energy corresponding to symmetric fission of $^{218}Ra$ calculated within the framework of the LSD model [31]. It reflects typical mutual layout of the ground state, the ridge, and the fission valley. For the present work this figure is of illustrative character.

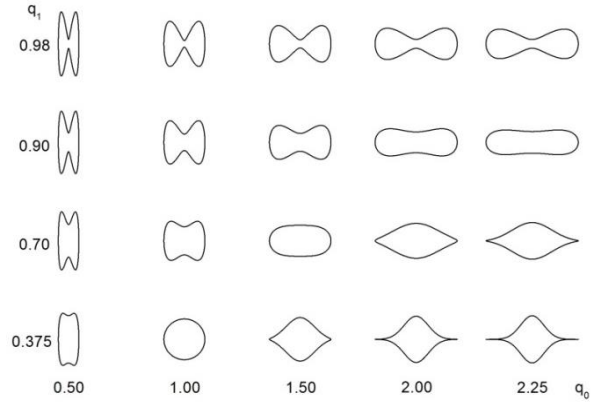

**Fig. 1.** The shapes of the fissioning nucleus for different values of the deformation parameters (only mirror-symmetric fission is considered).

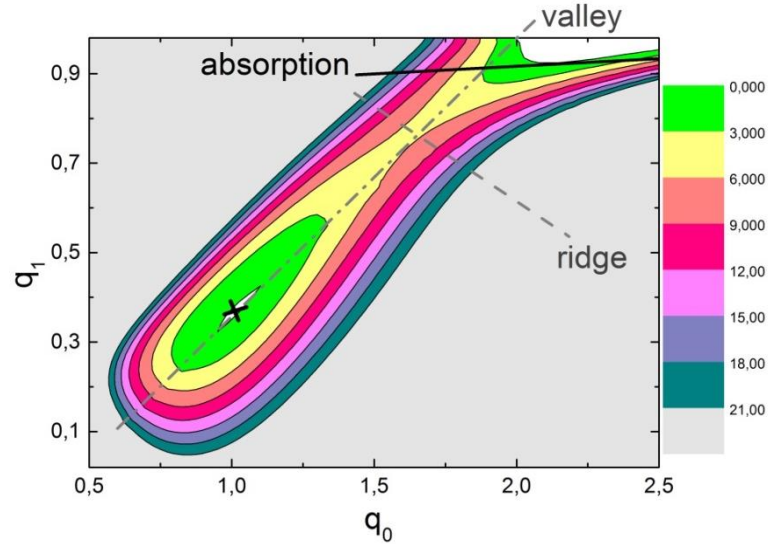

**Fig. 2.** The realistic potential energy landscape for $^{218}Ra$ along which the Brownian particle (fissioning nucleus) moves. The energy is calculated within the framework of the LSD model [31]. The ground state (cross), the ridge, the fission valley as well as the typical absorptive (scission) line are also shown.

At the initial moment of time the particle is located near the metastable state with the coordinates $q_{0c} = 1.00$, $q_{1c} = 0.375$ (see Figs. 1, 2). In the case of nuclear fission this corresponds to the situation when the nucleus has spherical shape. Because of thermal fluctuations, the Brownian particle can reach the barrier point with the coordinates $q_{0b}$, $q_{1b}$ or its vicinity, i.e. the ridge separating the quasistationary state from the valley in the upper-right corner of Fig. 2. The difference between the potential energies at $q_{0c}$, $q_{1c}$ and $q_{0b}$, $q_{1b}$, $U_b - U_c$, is called the barrier height in nuclear fission or the activation energy in chemical reactions. Henceforth we set $U_c = 0$. After reaching the ridge, the particle can return to the quasistationary state due to fluctuations or move further to the absorptive border due to the driving force. The absorptive border in nuclear fission corresponds to the scission line at which the nucleus separates quickly into two fragments.

In the present work for the potential energy we use the following analytical ansatz:

$$U(q_0, q_1) = U_{P2}(q_0) + C_1(q_1 - q_{1v})^2/2, \tag{1}$$

$$U_{P2}(q_0) = \begin{cases} C_0(q_0 - q_{0c})^2/2 & \text{at } q_0 < q_{0m}; \\ U_b - C_0(q_0 - q_{0b})^2/2 & \text{at } q_0 > q_{0m}. \end{cases} \tag{2}$$

Here

$$q_{1v}(q_0) = q_{1c} + \frac{q_{1b} - q_{1c}}{q_{0b} - q_{0c}}(q_0 - q_{0c}), \tag{3}$$

$$q_{0m} = (q_{0b} + q_{0c})/2, \tag{4}$$

$$C_0 = U_b/(q_{0b} - q_{0c})^2. \tag{5}$$

Equations (4) and (5) guarantee smooth connection of two pieces of the potential in Eq. (2) at $q_0 = q_{0m}$. The stiffness $C_1$ is taken to be $q_0$-dependent. We approximate this dependence using the formula [24]

$$C_1 = C_{1as}\{1 + w[1 + \exp\{(2q_0 - q_{0c} - q_{0b})/(q_{0c} - q_{0b})\}]^{-1}\}. \tag{6}$$

Here $C_{1as}$ is the asymptotic value of the stiffness ($C_1 \Rightarrow C_{1as}$ as $q_0 \Rightarrow \infty$) and $w$ controls the evolution of the width of the valley from the ground state to the saddle state and beyond.

We presume that the absorptive border (subscript '$a$') is represented by a straight line whose equation reads

$$q_{1a} = q_{1s} + k_0(q_0 - q_{0s}). \tag{7}$$

Two examples of the potential energy maps for $^{218}$Ra used for the dynamical modeling are presented in Fig. 3. They are calculated using Eqs. (1)-(5). As in Fig. 2 the ridge, the fission valley, and the typical absorptive (scission) lines are also shown. Panel a) corresponds to the case of $q_{1b} = q_{1c}$ which is referred to as "perpendicular valley". The cases $q_{1b} \neq q_{1c}$ will be referred to as "diagonal valley". One of such cases with $q_{1b} = 2q_{1c}$ is presented in Fig. 3b. One sees that the analytical potential presented in Fig. 3b reproduces main distinct features of the realistic potential shown in Fig. 2.

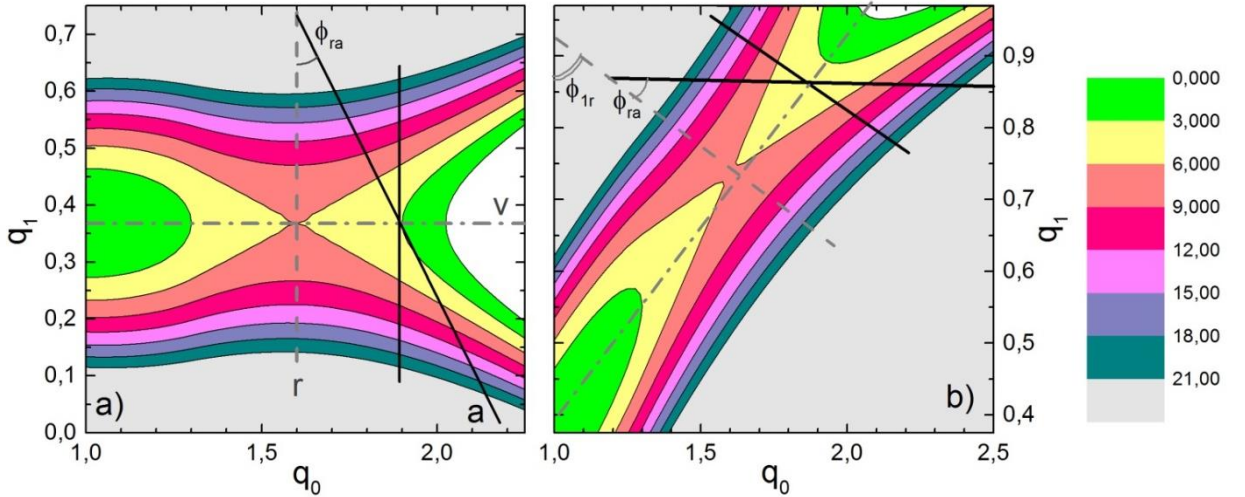

**Fig. 3.** Two analytical potential energy landscapes calculated using Eqs. (1)-(5) for $^{218}$Ra. As in Fig. 2 the ridge ('r'), the fission valley ('v'), and the typical absorptive (scission) lines ('a') are also shown. Panel a) corresponds to the case of $q_{1b} = q_{1c}$ ("perpendicular valley"); in panel b) $q_{1b} \neq q_{1c}$ ("diagonal valley").

The angles $\phi_{1r}$ and $\phi_{ra}$ shown in the figure are defined as follows. The angle of rotation of the ridge relatively to the $q_1$ axis is denoted as $\phi_{1r}$ whereas the angle of rotation of the absorptive border relatively to the ridge is denoted as $\phi_{ra}$ (the angles are positive in the case of the anti-clockwise rotation). The coefficient $k_0$ in Eq. (7) is related to these angles in the following way:

$$k_0 = -1/\tan(\phi_{ra} + \phi_{1r}). \tag{8}$$

The perpendicular valley in Fig. 3a corresponds to $\phi_{1r} = 0$; in Fig. 3b $\phi_{1r} = 0.56$ (note, that the vertical and horizontal scales are different in the figures).

## 2.2. Dynamical equations and corresponding decay rates

The time evolution of the dynamical variables of the Brownian particle is described by the stochastic differential equations (the Langevin equations). For the two-dimensional case these equations read (see, e.g. [3]):

$$\frac{dp_i}{dt} = -(\partial_i \mu_{jk}) p_j p_k / 2 - \eta_{ij} \mu_{jk} p_k + K_i + g_{ij} \Gamma_j, \tag{9}$$

$$\frac{dq_i}{dt} = \mu_{ik} p_k. \tag{10}$$

The symbol $\partial_i$ denotes the partial derivative with respect to $q_i$. The time evolution of the system is defined by the inverse inertia tensor $\mu_{jk}$, the friction tensor $\eta_{ij}$, the driving forces $K_i = -\partial_i U$, the random forces $g_{ij}\Gamma_j$. Equations (9), (10) represent a generalization of the classical dynamical Hamilton equations for the case when the mechanical system moves under the influence of the dissipative forces and fluctuations.

The amplitudes of the random forces are related to the temperature and the components of the friction tensor by the fluctuation-dissipation theorem:

$$g_{ik} g_{kj} = T \eta_{ij}. \tag{11}$$

In nuclear physics the temperature $T$ is measured in MeV as the potential energy, thus the Boltzmann constant is equal to 1. The random forces $\Gamma_i$ are taken to represent white noise

$$\langle \Gamma_i \rangle = 0, \tag{12}$$
$$\langle \Gamma_i(t_1) \Gamma_i(t_2) \rangle = 2\delta_{ij} \delta(t_1 - t_2). \tag{13}$$

In the discrete form corresponding to the Euler-Maruyama method [32] the Langevin equations read

$$p_i^{(n+1)} = p_i^{(n)} + \Delta p_i, \tag{14}$$
$$q_i^{(n+1)} = q_i^{(n)} + \Delta q_i, \tag{15}$$
$$\Delta p_i = -\{(\partial_i \mu_{jk}) p_j p_k / 2 + \eta_{ij} \mu_{jk} p_k - K_i\}\tau + g_{ij} b_j \sqrt{\tau}, \tag{16}$$
$$\Delta q_i = \mu_{ik}^{(n)} (p_k^{(n)} + p_k^{(n+1)})/2. \tag{17}$$

The superscripts represent two moments of time separated by the interval $\tau$, which is equal to the time step of numerical modeling. In the rhs of Eq. (16) all quantities correspond to the time moment $n\tau$. The random numbers $b_j$ entering the random forces have a Gaussian distribution with zero averages and variances equal to 2.

Although in the fission problem the inertia and friction tensors are deformation-dependent, in this work we ignore this dependence and consider the tensors to be diagonal:

$$\eta_{ij} = k_\eta \eta \delta_{ij} = k_\eta \hbar (n_0 A)^{4/3} r_0^4 \delta_{ij}, \tag{18}$$
$$\mu_{ij} = \frac{\delta_{ij}}{m} = (r_0^2 m_0 A^{5/3})^{-1} \delta_{ij}. \tag{19}$$

Here $n_0 = 0.17$ fm$^{-3}$ is the nucleon saturation density, $r_0 = 1.22$ fm, $m_0 = 0.01044$ MeV·zs$^2$/fm$^2$ is the nucleon mass, the dimensionless parameter $k_\eta$ allows to vary the friction strength, $m$ is the "mass" of the Brownian particle.

The definition of the time-dependent decay rate reads:

$$R_{at}(t) = -\frac{1}{\Pi(t)} \frac{d\Pi}{dt}. \tag{20}$$

Here $\Pi(t)$ is the probability that the metastable state has not yet decayed by time moment $t$.

Equations (14)-(17) describe the Markovian process, i.e. the memory effects are not taken into account. The equations are solved numerically using random numbers. The solution is actually a sequence of $N_{tot}$ trajectories all terminated not later than at the moment of time $t_D$. Some of those trajectories reach the absorptive border before $t_D$. The decay rate is calculated in this algorithm as follows

$$R_{at} = \frac{1}{N_{tot} - N_{at}} \frac{\Delta N_{at}}{\Delta t}. \tag{21}$$

Here $N_{at}$ is the number of Brownian particles (or stochastic trajectories) which have reached the absorptive border by the time moment $t$, $\Delta N_{at}$ is the number of particles which have reached the absorptive border during the time interval $\Delta t$. In order to find the $R_D$ we choose several bins beginning from the end of $R_{at}$-array (i.e. from the time moment $t_D$) and average the $R_{at}$ over these bins. This procedure and its errors are discussed in details in Ref. [33] (see also Appendix A).

Typical behavior of $R_{at}$ is shown in Fig. 4. Here and below for all the figures involving the decay rate the parameters of the calculations are collected in Table B1. After a transient stage, the decay rate reaches its quasistationary value $R_D$. Each time we checked whether the results of modeling did not depend upon the time step within the statistical errors. The value of $\tau$ typically was varied from 0.05 up to 0.20 zs. In order to calculate the quasistationary decay rate with the relative error 1% or smaller about 10 thousand fissioned trajectories were obtained.

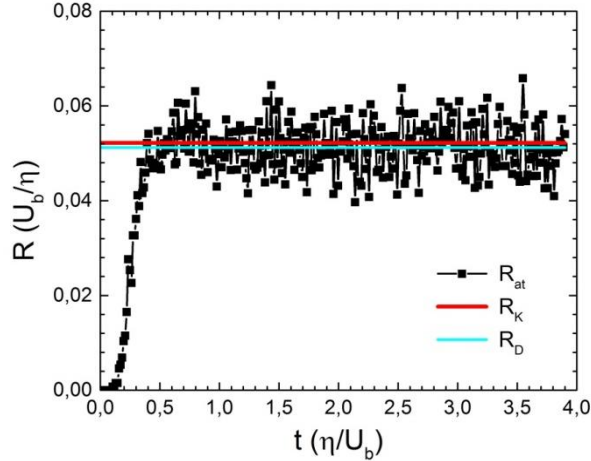

**Fig. 4.** The typical time dependence of the dynamical decay rate $R_{at}$ (oscillating line with boxes). The horizontal lines indicate the QDR ($R_D$) and the Kramers rate ($R_K$, see Eq. (22)). See parameters of the calculations in Table B1.

*2.3. Approximate analytical decay rate*

The generalization of the Kramers formula obtained in Refs. [15, 16] reads

$$R_K = \omega_K \left\{ \frac{|\det U_c''| \det m_b}{|\det U_b''| \det m_c} \right\}^{1/2} exp(-U_b/T). \qquad (22)$$

Here $\det U_c''$ ($\det U_b''$) is the determinant of the second derivatives of the potential energy at the quasistationary (saddle) point; $\det m_c$ ($\det m_b$) is the determinant of the inertia tensor at the quasistationary (saddle) point. The multiplier $\omega_K$ is the only positive root of the equation

$$\det \left\{ m_{ij} \omega_K^2 + \eta_{ij} \omega_K - \frac{\partial^2 U}{\partial q_i \partial q_j} \right\}_b = 0. \qquad (23)$$

Equation (22) is supposed to be valid if:
  (i) the potential barrier is high enough compare to the temperature;
  (ii) the absorptive border is far enough from the barrier (ridge);
  (iii) the quasistationary point is far enough from the barrier (ridge);
  (iv) the potential is represented well by the portions of parabolas near the quasistationary and saddle states.

## 3. Results

*3.1. Absorptive border far from the ridge*

Let us first study how the Kramers rate measures up against QDR when the absorptive border is far enough from the ridge. The simplest case is when the valley goes along $q_0$ coordinate (perpendicular valley). Then the only thing which can matter is the dependence of $C_1$ in Eq. (1) upon $q_0$ Fig. 5 illustrates this evolution for several values of $w$ (see Eq. (6)). The increase of $C_1$ with $q_0$ (the case of $w < 0$) should result in smaller values of the decay rate because the population at the ground state is larger (the ground state is wider). At $w > 0$ the situation turns to the opposite: a narrow ground state squeezes out the particles from it. These features were discussed in Ref. [22]. Resulting Kramers rate and the QDR are displayed in Fig. 6. We see ideal agreement between $R_K$ and $R_D$. The statistical errors of $R_D$ here and in all the relevant figures below are usually about 1% (except Appendix A).

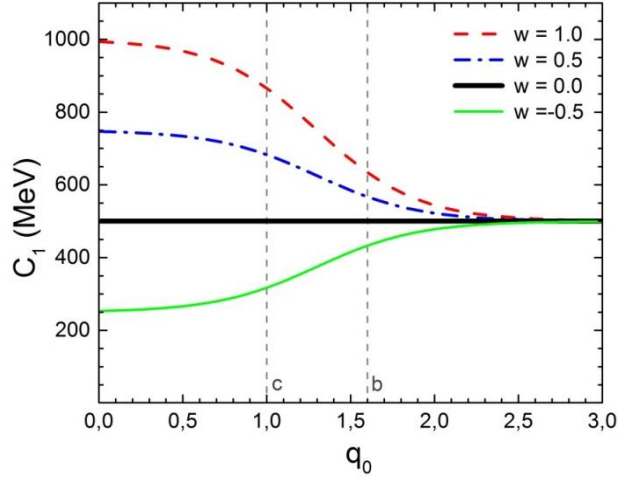

**Fig. 5.** Evolution of the stiffness of the valley with $q_0$ for several values of $w$ indicated in the figure. The vertical lines correspond to the ground state ('c') and to the barrier ('b').

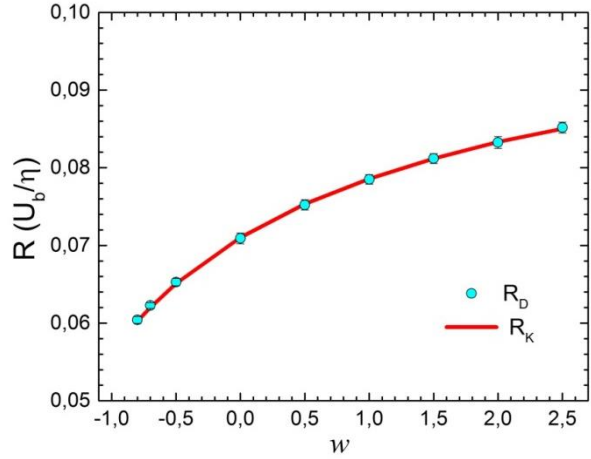

**Fig. 6.** Dependence of the decay rates upon the coefficient $w$ in Eq. (6). See parameters of the calculations in Table B1.

Now we consider the diagonal valleys like in Fig. 3b and $C_{1c} = C_{1b}$ (i.e. $w = 0$). Here $C_{1c} = C_1(q_{0c})$ and $C_{1b} = C_1(q_{0b})$. Thus the width of the valley does not depend on $q_0$ and we vary $q_{1b}$, i.e. rotate the valley in $(q_0, q_1)$ plane. Results of these calculations are shown in Fig. 7. We see that $R_D$ and $R_K$ agree perfectly with each other whereas both of them somewhat decrease as the valley rotates and the modes become coupled stronger. Careful analysis of the results shows that this decrease is due to the Kramers frequency $\omega_K$ in Eq. (22) solely. This frequency, plotted in Fig. 7 too, is the least physically clear ingredient of Eq. (22). We did not manage to find its physical interpretation in the literature. It looks like $\omega_K$ is related somehow to the distance between the quasistationary and barrier points: the larger this distance, the smaller $\omega_K$.

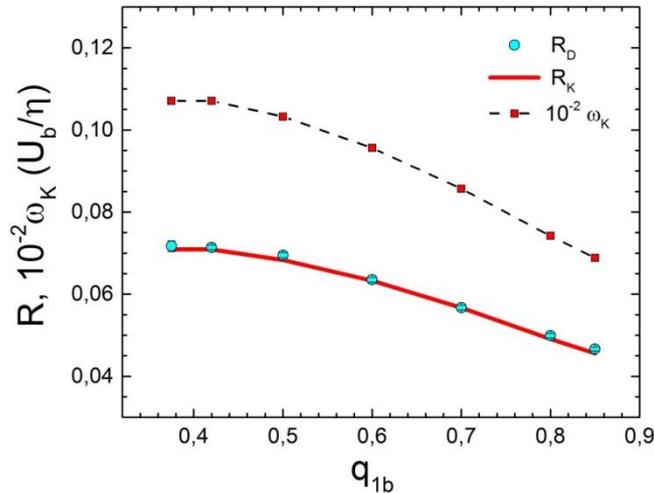

**Fig. 7.** The dependence of the decay rates $R_D$ and $R_K$ and of the Kramers frequency $\omega_K$ upon $q_{1b}$. See parameters of the calculations in Table B1.

Let us now vary $q_{1b}$ and $w$ simultaneously. The results are presented in Table 1 as the fractional differences

$$\xi = R_K/R_D - 1 \qquad (24)$$

with the statistical errors (both in percent) for different combinations of $C_{1b}/C_{1c}$ and $q_{1b}/q_{1c}$. The agreement within the statistical errors not exceeding 1% is observed. Thus we conclude that the orientation of the valley and the dependence of its stiffness upon the elongation coordinate $q_0$ cannot result in any inaccuracy of the Kramers rate.

**Table 1**
The fractional difference between $R_D$ and $R_K$, $\xi$ (see Eq. (24)), and its absolute statistical error $\Delta\xi$ (in brackets) for different combinations of $C_{1b}/C_{1c}$ and $q_{1b}/q_{1c}$ in the form $\xi(\Delta\xi)$. Both $\xi$ and $\Delta\xi$ are presented in percent; $\Delta\xi = \varepsilon_R$ (see Eq. (A2)).

| $C_{1b}/C_{1c}$ | $q_{1b}/q_{1c}$ | | | | |
|---|---|---|---|---|---|
| | 1.00 | 1.20 | 1.47 | 1.73 | 2.00 |
| 0.82 | 0.12 (0.78) | 0.40 (0.78) | 0.65 (0.57) | -0.59 (1.00) | -1.16 (0.98) |
| 0.89 | 0.16 (0.88) | 0.64 (0.84) | 0.40 (0.72) | 0.58 (0.99) | -0.36 (0.74) |
| 1.00 | 0.16 (0.97) | 0.73 (0.69) | 0.89 (0.73) | 0.65 (1.00) | 1.03 (0.76) |
| 1.19 | -0.14 (0.82) | 0.68 (0.64) | 0.39 (0.54) | 0.55 (0.82) | 0.26 (0.93) |
| 1.35 | -0.49 (0.50) | -0.05 (0.85) | -0.02 (0.92) | 0.11 (0.77) | -0.48 (0.60) |

*3.2. The influence of the absorptive border*

Let us now see how the Kramers rate measures up against QDR when the absorptive border moves closer to the saddle point being perpendicular to the valley. Here we again consider the case when the valley goes along $q_0$ coordinate (see Fig. 3a). Dependence of $R_D$ upon $q_{0s}$ (see Eq. (7)) at $\phi_{1r} = 0$ is presented in Fig. 8. The Kramers rate (thick horizontal line) agrees with the QDR (line with symbols) at $q_{0s} > 2.00$ within the statistical errors. As $q_{0s}$ decreases the QDR exceeds $R_K$ and at $q_{0s} \approx q_{0b}$ the $R_D$ approaches $2R_K$ (thin horizontal line). This effect was discussed earlier for the 1D case (see, e.g. [13, 15]). The physical reason is that those particles which overcome the barrier still have a chance to be returned back into the potential well due to fluctuations if they are not absorbed at the border. In the formula for the Kramers rate, the border is supposed to be very far from the barrier and all these re-scatterings are accounted for. As we move the absorptive border closer to the ridge some particles of the previously re-scattered are absorbed and contribute to the QDR. It was shown in [13, 15] that for the 1D case the Kramers rate was to be corrected to account for this effect and the correction factor reached 2 when $q_{0s} \approx q_{0b}$. Thus the dependence of $R_D/R_K$ as the function of $q_{0s}$ is well understood.

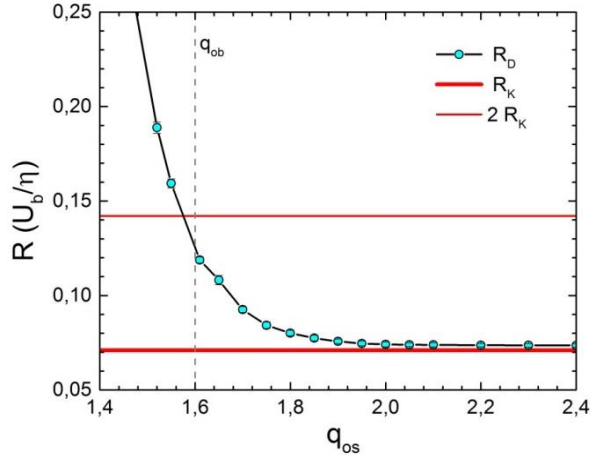

**Fig. 8.** The dependence of the decay rates upon $q_{0s}$ for the perpendicular valley. See parameters of the calculations in Table B1.

As the next step we study the effect of the slope of the absorptive border at two fixed values of $q_{0s}$. One of them ($q_{0s} = 2.0$) is large enough, i.e. it corresponds to that part of Fig. 8 where the sensitivity $R_D$ to $q_{0s}$ absents. The second value ($q_{0s} = 1.7$) is relatively close to the barrier. Results of such calculations are shown in Fig. 9. The angle between the ridge (perpendicular to the valley) and the absorptive border, $\phi_{ra}$, is used here as an argument. We rotate the absorptive border around the point with coordinates $q_{0s}$, $q_{1s}$ belonging to the valley (see Eq. (3)). For the symmetry reason, the dependence $R_D(\phi_{ra})$ should be even. One sees in Fig. 9 that this is the case indeed. Qualitatively we understand the increase of $R_D$ with $|\phi_{ra}|$ as follows. Due to the rotation of the absorptive border (line) one half of it is getting further from the ridge. This part of the line does not influence the QDR (see that part of Fig. 8 for which $q_{0s} > 2$). Another half of the line approaches the ridge and absorbs the particles which otherwise can be re-scattered. The net effect is the increase of $R_D$ with the increase of the slope. The following fast increase of $R_D$ as $|\phi_{ra}| => \pi/2$ happens because the absorptive border enters the vicinity of the ground state.

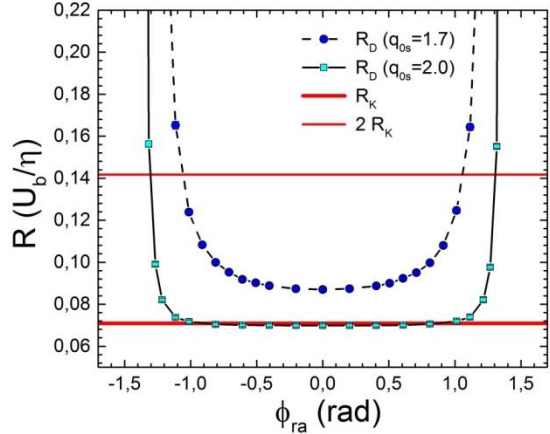

**Fig. 9.** The dependence of the decay rates upon the angle between the absorptive border and the ridge, $\phi_{ra}$, for the perpendicular valley for two values of $q_{0s}$ indicated in the figure. Other notations are as in the previous figure. See parameters of the calculations in Table B1.

Finally, let us consider a general situation: the diagonal potential like in Fig. 2 and Fig. 3b; the variable stiffness $C_1(q_0)$ with $w = -0.2$ (see Eq. (6)). As in Fig. 9 we rotate the absorptive border around a point in the valley. The minimum value of $R_D$ in these calculations is expected when this border is parallel to the ridge. The results are displayed in Fig. 10 for two cases: $q_{0s} = 1.7$ and $q_{0s} = 1.9$. The dependences $R_D(\phi_{ra})$ look similar to those of Fig. 9. However the graphs in Fig. 10 are somewhat asymmetric with respect to $\phi_{ra} = 0$. The increase of $R_D$ at the values of $|\phi_{ra}|$ departing from 0 is explained exactly the same way as in Fig. 9. The $q_0$-dependence of $C_1$ seems to not influence significantly the deviation of $R_K$ from $R_D$.

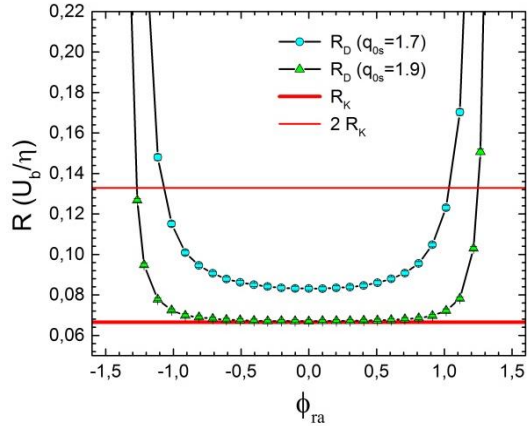

**Fig. 10.** The same as in Fig. 9 but for the diagonal valley and with the increasing stiffness ($w = -0.2$). See parameters of the calculations in Table B1.

*3.3. The influence of the friction strength*

In order to make our study more comprehensive we perform calculations for different values of $k_\eta$ (see Eq. (18)). Note, that the Kramers rate $R_K$ is not valid when friction is so weak that energy dissipation in one bouncing can be neglected. In the original Kramers work [11] for this case the following formula was obtained:

$$R_{K1} = \frac{\eta U_b}{mT} \, exp\left(-\frac{U_b}{T}\right). \qquad (25)$$

This formula was derived for the 1D motion and we are not aware about its multidimensional generalization. Therefore we compare in Fig. 11 results of numerical modeling with $R_{K1}$ and $R_K$. Results of this comparison tell us that $R_K$, in its domain of applicability, is valid for different values of the friction strength. Eq. (25) also does not contradict to the numerical decay rate for ballistic regime. We do not manage to obtain the values of $R_D$ for smaller $k_\eta$ because the Euler-Maruyama algorithm seems to break down.

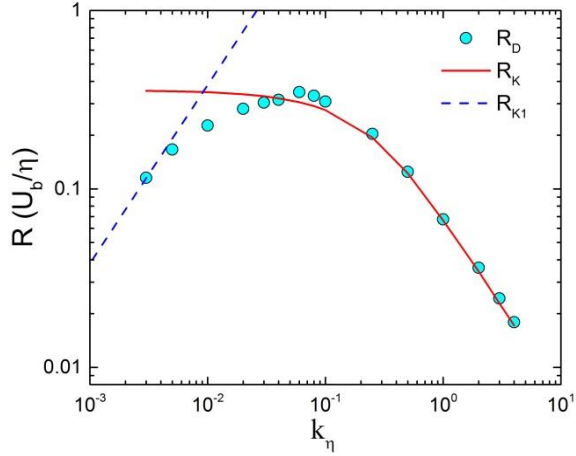

**Fig. 11.** The dependence of the decay rates upon the strength of friction $k_\eta$ (see Eq. (18)). Straight dash line indicates the 1D Kramers formula (25), other notations are as in Figs. 8, 9, 10. See parameters of the calculations in Table B1.

**4. Conclusions**

We have studied the accuracy of the Kramers approximate formula (22) for the thermal decay rate of the metastable state for the two-dimensional potential pocket. For this aim we have modeled the decay process solving numerically the Langevin-type stochastic equations and building the quasistationary rate on the basis of this solution. The modeling has been performed both for the coupled and uncoupled modes. The potential has been composed of the second-order parabolas thus the anharmonic effects have been excluded.

It has been shown that the Kramers rate is in agreement with the quasistationary rate within the statistical errors (1%) provided the absorptive border is far enough from the potential ridge restricting the metastable state. This result holds for different orientations and shapes of the valley leading from the metastable state. As the absorptive border (or its part) gets closer to the ridge, the Kramers formula underestimates the numerical quasistationary rate. The difference reaches approximately the factor of 2 when the absorptive border coincides with the ridge. These results were obtained earlier for the 1D overdamped case. Thus, it turns out that the influence of the absorptive border on the accuracy of the Kramers formula does not depend upon the dimensionality of the modeling and upon the strength of friction in the diffusive regime.

Note, that according to [15, 17] our results can be alternated by an anharmonicity of the potential. The non-diagonal terms of the inertia and friction tensors (as well as the typical for the fission problem coordinate dependence of those tensors) could influence the agreement between the QDR and the Weidenmüller's formula for $R_K$. We plan to address this problem in near future.

**Acknowledgments**

M. V. C. is grateful to the Dmitry Zimin ''Dynasty'' Foundation for the financial support.

**Appendix A**

Numerical modeling of the random motion of a Brownian particle is based on a pseudo-random number generator. Results of this modeling is a sequence of $N_{tot}$ trajectories each terminated not later than at $t_D$. The trajectories which have reached the absorptive border during this time lapse contribute to the decay rate $R_{at}$ calculating according to Eq. (21) (see also Fig. 4). The time dependence of $R_{at}$ can be separated into a nonlinear transition part and a quasistationary part although significant fluctuations may be present. We are interested in the quasistationary value of the rate, $R_D$, which is calculated as an average value of $R_{at}$ over the quasistationary part

$$R_D = \frac{1}{k} \sum_{j=L-k}^{L} R_{at}(t_j). \tag{A1}$$

Here $L = t_D/\Delta t$ is the total number of bins, $\Delta t$ is the bin width, $k$ is the number of bins used for finding $R_D$, the initial $(L-k)$ bins are disregarded. The aim of this Appendix is to show that the value of $R_D$ does not depend (within the statistical errors) upon $\Delta t$, $k$, and $N_{tot}$. The statistical error of $R_D$ reads

$$\varepsilon_R = \frac{1}{R_D \sqrt{k}} \sqrt{\frac{1}{k-1} \sum_{j=L-k}^{L} \left(R(t_j) - R_D\right)^2}. \quad (A2)$$

It must decrease as $\sqrt{k}$.

Checking these properties of $R_D$ and $\varepsilon_R$ is of importance due to the well-known periodic character of any pseudo-random number generator. The larger is the number of trajectories (and presumably the smaller is $\varepsilon_R$) the larger is the probability of catching this periodicity.

It is convenient to analyze not the QDR itself but its fractional difference from the Kramers rate $\xi$ defined by Eq. (24). This fractional difference and its absolute error which is equal to $\varepsilon_R$ are shown in Fig. A1. In the upper panels we see that as the interval of time-averaging $(k\Delta t)$ increases, $\xi$ somewhat decreases and then stays stable fluctuating within 1%. When $(k\Delta t)$ exceeds the duration of the quasistationary stage, $\xi$ begins to grow sharply, indicating that $R_D$ gets smaller. This is expected because the transient stage (where $R_{at}$ significantly smaller than the QDR) becomes involved in the calculation. In the lower panels of Fig. A1 the relative error of $R_D$, $\varepsilon_R$, is shown. It evolves with $(k\Delta t)$ according to $k^{-1/2}$ law. As the transient stage starts to be involved, $\varepsilon_R$ sharply increases. The curves without symbols in lower panels correspond to $(k\Delta t)^{-1/2}$ dependence and has been adjusted to $\varepsilon_R$ at some intermediate points. One sees that all these conclusions are stable with respect to the value of $\Delta t$ (compare left and right columns).

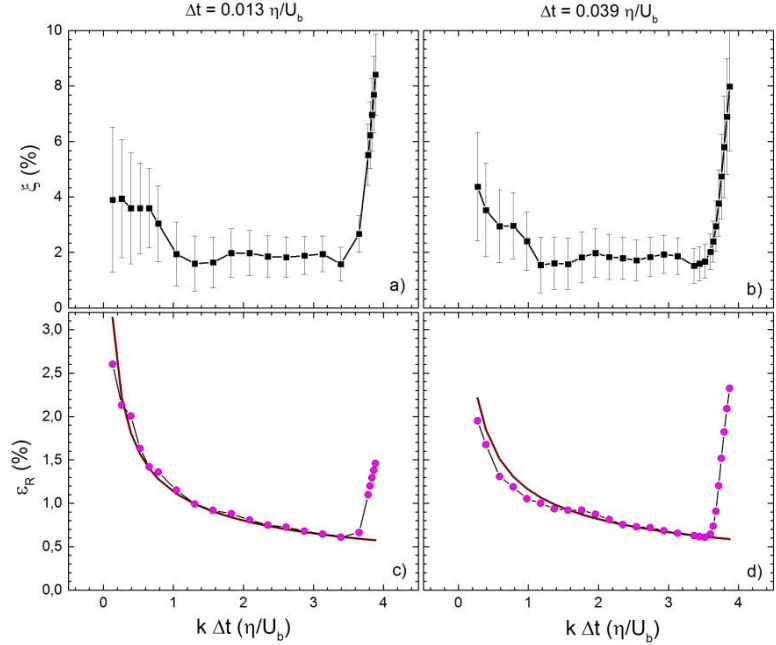

**Fig. A1.** The fractional difference $\xi$ defined by Eq. (24) (panels a and b) and its error $\varepsilon_R$ (panels c and d) versus the interval of time-averaging $(k\Delta t)$ for two values of $\Delta t$ (left and right columns). The curves without symbols in panels c and d represent the $(k\Delta t)^{-1/2}$ dependence and has been adjusted to $\varepsilon_R$ at some intermediate points. See parameters of the calculations in Table B1.

In Fig. A2 we present the dependence of $R_D$ and $\varepsilon_R$ upon the number of trajectories. It is seen that as this number increases the $R_D$ approaches its constant value and the relative error decrease like $N_{tot}^{-1/2}$ as it is expected. Note that even at very small number of trajectories ($N_{tot} = 100 \div 500$) when the number of useful trajectories reaching the absorptive border is extremely small, our algorithm produces $R_D$ which is only 20-30% away from the more correct value corresponding to $2 \cdot 10^5$ trajectories. Thus using this algorithm one can obtain rough estimate of $R_D$ very quickly.

Thus we conclude that all results of the present paper are valid and no evidence of generator periodicity is seen.

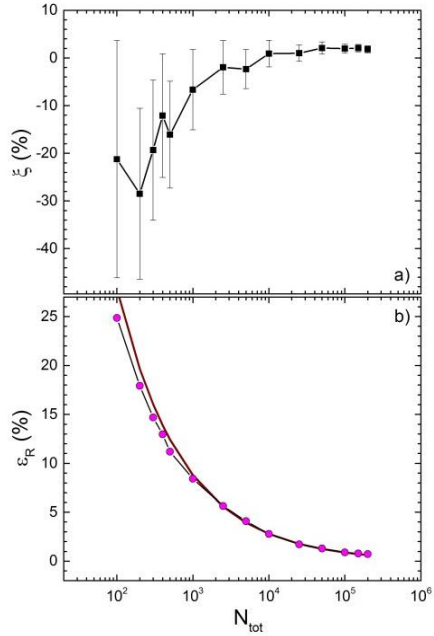

**Fig. A2.** Same as in Fig. A1 but versus the total number of trajectories $N_{tot}$. The curve without symbols in panel b represents the $N_{tot}^{-1/2}$ dependence and has been adjusted to $\varepsilon_R$ at some intermediate points. See parameters of the calculations in Table B1.

## Appendix B

**Table B1**
Parameters of the modeling for the figures involving the decay rate.
$U_b = 6.0$ MeV, $\eta = 460$ MeV zs, $T = 1.89$ MeV, $q_{0b} = 1.6$, $C_{1as} = 500$ MeV.

| Figure | $\tau$, zs | $w$ | $q_{1b}$ | $q_{0s}$ | $\phi_{1r}$ | $\phi_{ra}$ | $k_\eta$ |
|---|---|---|---|---|---|---|---|
| Fig. 4 | 0.10 | 0.0 | 0.700 | 2.20 | 0.0 | 0.0 | 1 |
| Fig. 6 | 0.05-0.20 | variable | 0.375 | 2.50 | 0.0 | 0.0 | 1 |
| Fig. 7 | 0.10 | 0.0 | variable | 2.00 | 0.0 | 0.0 | 1 |
| Fig. 8 | 0.10 | 0.0 | 0.375 | variable | 0.0 | 0.0 | 1 |
| Fig. 9 | 0.10 | 0.0 | 0.375 | 2.00 and 1.70 | 0.0 | variable | 1 |
| Fig. 10 | 0.10 | -0.20 | 0.500 | 1.90 and 1.70 | 0.205 | variable | 1 |
| Fig. 11 | 0.01-0.10 | -0.20 | 0.500 | 1.90 | 0.205 | 0.5 | variable |
| Figs. A1,A2 | 0.10 | 0.0 | 0.700 | 2.20 | 0.0 | 0.0 | 1 |